\RequirePackage{fix-cm}
\documentclass[twocolumn,epjc3]{svjour3}
\smartqed  
\RequirePackage{graphicx}
\journalname{Eur. Phys. J. C}
\begin{document}

\title{Periodicity and area spectrum of black holes}

\author{Xiao-Xiong Zeng \thanksref{e1, addr1}
        \and Xian-Ming Liu
        \thanksref{e2, addr2, addr1}  \and
Wen-Biao Liu(corresponding author)
        \thanksref{e3, addr1}}

\thankstext{e1}{e-mail: xxzeng@mail.bnu.edu.cn}
\thankstext{e2}{e-mail: liuxianming1980@163.com}
\thankstext{e3}{e-mail: wbliu@bnu.edu.cn}
\institute{Department of Physics, Institute of Theoretical Physics,
Beijing Normal University, Beijing 100875, China \label{addr1}
          \and
         Department of Physics, Hubei University for Nationalities, Enshi
445000, Hubei, China \label{addr2} }



\date{Received: date / Accepted: date}

\maketitle

\begin{abstract}
The recent speculation of Maggiore that the periodicity of a black
hole may be the origin of the area quantization law is confirmed. We
exclusively utilize the period of motion of an outgoing wave, which
is shown to be related to the vibrational frequency of the perturbed
black hole, to quantize the horizon areas of a Schwarzschild black
hole and a Kerr black hole. It is shown that the equally spaced area
spectrum for both cases takes the same form and the spacing is the
same as that obtained through the quasinormal mode frequencies.
Particularly, for a Kerr black hole, the small angular momentum
assumption,  which is necessary from the perspective of quasinormal
mode, is not employed as the general area spacing is reproduced.

\keywords{area spectrum;\ periodicity;\ black hole} \
 \PACS{04.70.Dy, \ 04.70.-s}
\end{abstract}
\section{Introduction}
In recent several years, investigation on quasinormal modes of black
holes has attracted more and more attention of astrophysicists and
theoretical physicists. It is believed  that quasinormal modes of
black holes can provide not only a way to find black holes by
detecting their fingerprints \cite {Kokkotas1999} but also a tool to
check the complex quantum gravity theory \cite {Konoplya83}. In
addition, it was shown that the quasinormal modes can also provide a
method to quantize the horizon area of black holes  on the basis of
Bohr's correspondence principle, which states that transition
frequencies at large quantum numbers are equal to classical
oscillation frequencies.  According to the initial result of Hod
\cite {Hod1998}, the real part of the quasinormal mode frequency is
responsible for the  area spectrum of black holes. Based on the
quasinormal modes of Schwarzschild black hole in large $n$ limit
\cite {Nollert16,Leaver402}, namely
 \begin{equation}
 8\pi M \omega_n=\ln3+2 \pi i (n+\frac{1}{2})+O(n^{-1/2}), \label{1}
 \end{equation}
Hod got the quantized horizon area   $\Delta A=32 \pi M \delta M=4
\ln3 l_{p}^2 $. This result was confirmed by  Kunstatter
\cite{Kunstatter2003} later using an adiabatic invariant
  \begin{equation}
\int\frac{dM}{\omega}\equiv A\simeq n \hbar \ \ \ \ as  \ \ \ \ n  \
\rightarrow \infty, \label{2}
\end{equation}
where  $\omega$  and $M$ are quasinormal mode frequency and black
hole mass, respectively. Actually, Kunstatter's idea stems from the
analogy with classical harmonic oscillator.  It is found that the
action integral of the form $A =\oint p dq $ for a quasiperiodic
system is an adiabatic invariant in analytical mechanics.  For a
one-dimensional harmonic oscillator with Hamiltonian
$H=\frac{p^2}{2m}+\frac{m\omega_c^2q^2}{2}$, the adiabatic invariant
was shown to be $A= E/\omega_c $\cite{Wells2007}.
 As the classical
vibrational frequency $\omega_c$ and system energy $ E $  are
treated as the quasinormal mode frequencies $\omega$ and black hole
mass $M$ in large $n$ limit, one can get Eq.(\ref{2}) immediately.
 In 2007, Maggiore \cite{Maggiore2008} elucidated that the treatment of Hod and Kunstatter  should be reexamined.
 He proved that the proper frequency of the equivalent harmonic oscillator,
which is interpreted as the quasinormal mode frequency, contains
contributions of both the real part $\omega_R$ and imaginary part
$\omega_I$ in  high damping limit. More importantly, he found that
the imaginary part  rather than the real part is dominant for the
highly excited quasinormal modes. Therefore, one should use the
imaginary part of quasinormal mode frequencies to study the  area
spectrum of black holes. For the case of Schwarzschild black hole,
the quantized horizon area  was shown to be $ \Delta A=8\pi
l_{p}^2$, which is obviously different from the one obtained by Hod
and Kunstatter. Another important speculation in Maggiore's work is
that the periodicity  of a black hole  in  Euclidean time may be the
origin of the area quantization. In fact, this inference can be got
from Eq.(1) directly. Considering the contributions of real part and
imaginary part of quasinormal modes,  the transition frequency in
large $n$ limit can be expressed as
$\omega=\omega_n-\omega_{n-1}=2\pi/T_{BH}$. It is well known that
for any background space time with a horizon in Kruskal coordinates,
the period with respect to Euclidean time takes the form as
$T={2\pi}/{\kappa_h}$  \cite{Gibbons1978}, here $\kappa_h$ is the
surface gravity  that relates to the temperature of the black hole
with the relation $T_{BH}=(\hbar \kappa_h)/2\pi$. Hence one can see
immediately that the transition frequency is related to the period
of  the  black hole  in  Euclidean time.

In fact,  horizon area of black holes is quantized in units of
$l_{p}^2$ was proposed firstly by Bekenstein
\cite{Bekenstein467,Bekensteingr-qc/9710076,Bekensteingr-qc/9808028}.
He found that the horizon area of a non-extremal black hole is
adiabatic invariant classically. According to the Ehrenfest
principle, any classical adiabatic invariant corresponds to a
quantum entity with discrete spectrum, Bekenstein conjectured that
the horizon area of a non-extremal quantum black hole has a discrete
eigenvalue spectrum. Based on Christodoulou's point particle model
\cite{Christodoulou1970}, Bekenstein found that the smallest
possible increase in horizon area of a non-extremal black hole is $
\Delta A=8\pi l_{p}^2$. Obviously, Maggiore's result
\cite{Maggiore2008} is consistent with it.

Following the work of Maggiore, the area spectrum of a lot of black
holes has been investigated.  Exploiting models proposed by Hod and
Kunstatter, Vagenas \cite{Vagenas0811} gave the area spectrum of a
Kerr black hole. It was found that there is a logarithmic term when
Kunstatter's model is employed. Soon after, Medved \cite {Medved25}
argued that the small angular momentum limit should be imposed in
Maggiore's work. Because the Bohr--Sommerfeld quantization condition
is actually an implication that the black hole should be far away
from the extremal case, the two methods used by Maggiore were found
to coincide with each other and both gave an equally spaced area
spectrum. Later on, area spectrum of charged black holes
\cite{Wei2010,Lopez-Ortega} was studied from the viewpoint of
quasinormal modes. It was found that the small charge limit should
be imposed in order to get the general area spacing $8\pi l_{p}^2$.
Until now, the area spectrum in de Sitter space times \cite
{li676,Chen69}, non-Einstein gravity \cite
{Kothawala104018,Majhi49,Banerjee279,Wei03,jiang2010}, and other
background space times \cite
{Kwon27,Kwon27165011,Kwon282011,Daghigh26,Wei0903} have been
discussed extensively via quasinormal modes analysis.

Now, we intend to give the area spectrum of black holes  without
using the quasinormal mode frequencies. In fact, there have been
some similar ideas before. In Ref.\cite{Banerjee2010}, it was found
that area spectrum of black holes can be obtained by computing the
average squared energy of the outgoing wave in the view of quantum
tunneling. In Ref.\cite{Ropotenko80}, Ropotenko stated that
quantization of the angular momentum component with commutation
relation in quantum mechanics can be used to quantize the horizon
area  of black holes. Recently, Majhi and Vagenas \cite{Majhi2011}
elucidated that an adiabatic invariant quantity, $\int {p_i dq_i}$,
can also be used to quantize the horizon area. Here we would like to
employ the periodicity of outgoing wave to obtain area spectrum of
black holes.  For a perturbed black hole, the outgoing wave performs
periodic motion outside the horizon and the period of motion is
related to the frequency of outgoing wave. It is well known that the
gravity system in Kruskal coordinates is periodic with respect to
Euclidean time. Particles' motion in this periodic gravity system
also has a period, which has been shown to be the inverse Hawking
temperature \cite{Gibbons1978}.  Therefore the concrete formulism of
frequency of outgoing wave  can be  given by the inverse Hawking
temperature \footnote[1] {We  thank  Prof. Elias  C. Vagenas for
pointing out that the period of motion of outgoing wave has the same
value as that of gravity system in Kruskal coordinates with respect
to the Euclidean time.}. In this case, the area spectrum  of black
holes can be reproduced directly with the proposal of Hod.

In Sec.2, the area spectrum of the  Schwarzschild black hole is
given with the proposal of Hod. We get the concrete value of
vibration frequency by equaling the motion period of outgoing wave
to the period of gravity system in the form of Euclidean time. In
Sec.3, our idea is extended to a Kerr black hole. The general area
spacing $8\pi l_{p}^2$ is obtained without any assumptions. Finally,
some conclusion and a discussion is given in Sec.4.

\section{Area spectrum of a Schwarzschild black hole}

The line element of a Schwarzschild black hole is
\begin{equation}
 ds^{2}=-f(r)dt^{2}+f^{-1}(r)dr^{2}+r^{2}d\theta ^{2}+r^{2}\sin^{2}\theta d\varphi^{2},\label{3}
\end{equation}
where%
\begin{equation}
 f(r)=1-\frac{2M}{r}.  \label{4}
\end{equation}%
 The location of the black hole horizon, namely $r_h$, is determined by $f(r_h)=0$.
 The Hawking temperature of this spacetime is $T_{BH}=\frac{\hbar \kappa_h}{2\pi}=\frac{\hbar}{8\pi M}$.
 Substituting Eq.(\ref{3}) into the Klein--Gordon equation
\begin{equation}
g^{\mu\nu}\partial _{\mu}\partial
_{\nu}\Phi-\frac{m^2}{\hbar^2}\Phi=0, \label{5}
\end{equation}
and adopting the wave equation ansatz $ \Phi=\frac{1}{{4\pi
\omega}^{1/2}}\frac{1}{r}$\\$R_{\omega}(r,t)Y_{l,m}(\theta,\phi)$\cite{Damora14}
for the scalar field, we can get the solution of wave function. On
the other hand, we  can also obtain the wave function  by the
Hamilton--Jacobi equation \footnote[2]{In fact, the Hamilton--Jacobi
equation can be obtained from the Klein--Gordon equation after
inserting Eq.(\ref{7}) into it and keeping the items only in order
of $\hbar$ \cite{Kerner25,Srinivasan204007}.}
\begin{equation}
g^{\mu\nu}\partial_{\mu}S\partial_{\nu}S+m^2=0,\label{6}
\end{equation}
where the action $S$ and the wave function $\Phi$ have the relation
\begin{equation}
\Phi=\exp[\frac{i}{\hbar}S(t,r,\theta, \phi)].\label{7}
\end{equation}
Now, we will concentrate on using the Hamilton--Jacobi equation to
find the wave function. For the spherically symmetric Schwarzschild
black hole, the action $S$ can be decomposed as \cite
{Kerner25,Srinivasan204007}
\begin{equation}
S(t,r,\theta, \phi)=-E t+W(r)+J(\theta,\phi),\label{8}
\end{equation}
where $E=-\partial S/ \partial t$ stands for the energy of emitted
particles observed at infinity. Near the horizon, $J$ vanishes and
$W$ can be solved as \footnote[3]{This value is used to investigate
Hawking tunneling radiation usually \cite {Kerner25} under the
relation $\Gamma=\exp[-2 Im S]$, where $\Gamma$ is the tunneling
probability, because only this part contributes to the imaginary
part of action.} \cite{Kerner25,Srinivasan204007}
\begin{equation}
W(r)=\frac{i \pi E}{f^{\prime}(r_h)},\label{9}
\end{equation}
where we only consider the outgoing wave near the horizon.  In this
case, it is obvious that the wave function $\Phi$ outside the
horizon can be expressed as the form
\begin{equation}
\Phi=\exp[-\frac{i}{\hbar}E t]\psi(r_h),\label{10}
\end{equation}
 where $\psi(r_h)=\exp[-\frac{ \pi E}{\hbar f^{\prime}(r_h)}]$.  From   Eq.(\ref{10}), we can see that $\Phi$ is a  periodic function
 with period
 \begin{equation}
T=\frac{2 \pi \hbar}{E}.\label{11}
\end{equation}
Taking into account the relation $E=\hbar \omega$, we find
\begin{equation}
T=\frac{2 \pi}{\omega}.\label{12}
\end{equation}
Apparently, the  frequency is related to the period of the outgoing
wave near the horizon. According to Hod's speculation, the change of
horizon area of the Schwarzschild black hole takes the form $\Delta
A= 32\pi \hbar M \omega $, where $\omega$ is the frequency of the
perturbed black hole. Here, $\omega$ can be regarded as the
frequency of outgoing wave, so the change of the horizon area can be
expressed as
\begin{equation} \Delta A= 32\pi M\hbar \times \frac{2\pi}{T} ,
\label{a1}
\end{equation}
where we have used Eq.(\ref{12}). It is well known that in Kruskal
coordinates, the gravity system  is a periodic system with respect
to the Euclidean time. Particles moving in this background hence
also have a  period\footnote[4]{Based on the Temperature Green
Function \cite{Gibbons1978}, it has been shown that this period is
the geometric origin of Hawking thermal radiation.}, which should be
the same as that of the periodic gravity system, namely
\cite{Gibbons1978}
\begin{equation}
T=\frac{2\pi}{\kappa_h}=\frac{\hbar}{T_{BH}}. \label{13}
\end{equation}

In Ref.\cite{Majhi2011}, this period has been used to study the
action of particle motion to find area spectrum of black holes. Here
we will also resort to this skill. Substituting Eq.(\ref{13}) into
Eq.(\ref{a1}), we can get immediately
\begin{equation}
\Delta A=8\pi l_{p}^2.  \label{14}
\end{equation}
Evidently, the area spectrum of a Schwarzschild black hole is
quantized with spacing $8 \pi l_{p}^2$. This is consistent with that
obtained by Maggiore \cite{Maggiore2008} from the viewpoint of
quasinormal mode.

\section{Area spectrum of a Kerr black hole}
The line element of a Kerr black hole is
\begin{eqnarray}  \label{1}
ds^{2}&=&-(1-\frac{2Mr}{\rho^2})dt_k^2+\frac{\rho^2}{\Delta}dr^2+
\rho^2d\theta^2+[(r^2+a^2)\sin^2\theta \nonumber\\
&+&\frac{2Mra^2\sin^4\theta}{\rho^2%
}]d\phi^2 -\frac{4Mra\sin^2\theta}{\rho^2}dt_kd\phi, \label{19}
\end{eqnarray}
where $\Delta =r^2-2Mr+a^{2},\quad \rho ^{2}=r^{2}+a^{2}\cos
^{2}\theta $. The outer(event) horizon and inner horizon of the Kerr
space time can be expressed as $r_{\pm }=M\pm (M^{2}-a^{2})^{1/2}$,
where the parameters $M$ and $a=J/M$ represent the mass, and the
angular momentum per unit mass, respectively. The Hawking
temperature and horizon area are
\begin{equation}
T_{BH}=\frac{\kappa_h}{2\pi}=\frac{\sqrt{M^{4}-J^{2}}}{4\pi
M\left(M^{2}+\sqrt{M^{4}-J^{2}}\right)} \hspace{1ex}, \label{20}
\end{equation}
\begin{equation}
A=\int(r_h^2+a^2)\sin \theta d\theta d\phi
=8\pi\left(M^{2}+\sqrt{M^{4}-J^{2}}\right).\label{21}
\end{equation}
For a Kerr black hole, there is an ergosphere between the outer
horizon and infinite redshift surface. To avoid the dragging effect,
usually one should perform the so-called dragging coordinate
transformation, where the dragging angular velocity at the event
horizon is defined as
\begin{equation}
\Omega_h=\frac{J}{2M\left(M^{2}+\sqrt{M^{4}-J^{2}}\right)}.\label{22}
\end{equation}
In this case, Eq.(\ref{1}) takes the form as
\begin{eqnarray}
ds^{2}&=&-\frac{\rho^2\Delta}{(r^2+a^2)-\Delta a^2\sin^2 \theta}dt_d^2+\frac{%
\rho^2}{\Delta}dr^2+ \rho^2d\theta^2  \nonumber\\
&=& -F(r)dt_d^2+\frac{1}{G(r)}dr^2+ H(r)d\theta^2. \label{23}
\end{eqnarray}
In the dragging coordinate frame, the action $S$ can be decomposed
as \cite{Chen665}
\begin{equation}
S(t,r,\theta)=-(E -m \Omega_h)t+W(r)+\Theta(\theta),\label{24}
\end{equation}
where $E$ is the energy of the emitted particle measured by the
observer at the infinity, $m$ denotes the angular quantum number
about $\phi$. Incorporating Eqs.(\ref{5}) and (\ref{23}), we find
near the horizon $\Theta$ vanishes and $W$ can be solved as
\begin{equation}
W(r)=\frac{i \pi (E -m
\Omega_h)}{\sqrt{F^{\prime}(r_h)G^{\prime}(r_h)}}, \label{25}
\end{equation}
where we also only consider the outgoing wave. So the wave function
$\Phi$ can be expressed as the form
\begin{equation}
\Phi=\exp[-\frac{i}{\hbar}(E -m \Omega_h) t]\psi(r_h). \label{26}
\end{equation}
Obviously, $\Phi$ is a periodic function with the period
\begin{equation}
T=\frac{2 \pi}{(\omega -(m \Omega_h)/\hbar)}. \label{27}
\end{equation}
Based on Eq.(\ref{21}), the change of horizon area of a Kerr black
hole can be written as
\begin{eqnarray}
\Delta A&=&8\pi [\frac{2M(\sqrt{M^{4}-J^{2}}+M^2)dM-2JdJ}{2\sqrt{M^{4}-J^{2}}}] \nonumber\\
&=&8\pi(\frac{dM}{2\pi
T_{BH}}-\frac{JdJ}{\sqrt{M^{4}-J^{2}}})].\label{28}
\end{eqnarray}
This treatment is different from that from the viewpoint of
quasinormal mode frequency \cite {Vagenas0811}, where only the
change of the mass $M$ is considered. In Ref.\cite {Vagenas0811},
the condition $\omega_I \gg \omega_R $ implies the imposition $M^2
\gg J$.

Combining Eq.(\ref{13}) and  Eq.(\ref{27}), we obtain
 \begin{equation}
dM=\hbar \omega=m\Omega_h+2\pi T_{BH}. \label{29}
\end{equation}
Substituting Eq.(\ref{29}) into  Eq.(\ref{28}) leads to
\begin{equation}
\Delta A=8\pi (1+\frac{m \Omega_h}{2\pi
T_{BH}}-\frac{JdJ}{\sqrt{M^{4}-J^{2}}}). \label{30}
\end{equation}
Simplifying it with Eq.(\ref{20}) and  Eq.(\ref{22}), we get
\begin{equation}
\Delta A=8\pi l_{p}^2. \label{31}
\end{equation}
Obviously, the equally spaced area spectrum of a Kerr black hole
produced here is consistent with that obtained from the viewpoint of
quasinormal mode \cite {Vagenas0811}. However, the small angular
momentum limit, which is necessary from the perspective of
quasinormal mode analysis, is not necessary as the general area gap
$8\pi\l_p^2$ is obtained.

\section{Conclusions and discussions}
To conclude, a new  scheme to quantize the horizon area of a black
hole was proposed. It was found that the period of the gravity
system with respect to the Euclidean time  can determine the area
spectrum of black holes exclusively. Obviously, our result confirmed
the speculation of Maggiore that the periodicity of a black hole may
be the origin of the area quantization. Here, the quasinormal mode
frequency is not used  and there is also no confusion on whether the
real part or imaginary part is responsible for the area spectrum. It
is more convenient and simple. If one wants to study area spectrum
of more complicated background space times from the viewpoint of
quasinormal mode to confirm the conjecture of Bekenstein that the
area spectrum is independent of black hole parameters, there are
some difficulties to find the quasinormal modes frequency
mathematically which can produce the area spectrum. Even for a Kerr
black hole, there are some controversies on the different formalisms
\cite {Vagenas0811}. In addition, our treatment is also different
from the quantum tunneling method \cite{Banerjee2010}. For the
quantum tunneling method, the perturbed frequency is obtained
through computing the average squared energy of the outgoing wave.
Here, the frequency is determined by the period of motion of
outgoing wave. Another important difference in the two methods is
the area spacing. For the quantum tunneling method,
  the spacing of the equally spaced area spectrum is $\Delta A= 4\l_p^2$, which is  smaller than the general value $8\pi\l_p^2$ obviously.
 The reason for this difference is that the  tunneling mechanism in the quantum tunneling method is similar to the Schwinger mechanism \cite{Srinivasan204007,Kim1095}, which will contribute to the quantum  of the quantized horizon area \cite{Hod204014}.

We have found that, for a perturbed black hole, the frequency is
determined by the period of the motion of an outgoing wave. It is
well known that the gravity system in the form of Euclidean time is
periodic. The concrete formalism of period, and further the
frequency of outgoing wave, can be given because of the same
periodicity. For any space times, we can get the quantized horizon
area easily. As examples of application, the area spectrum of a
Schwarzschild black hole and a Kerr black hole were studied in this
paper, and we found that our results are consistent with those
obtained from the viewpoint of quasinormal mode frequencies. We only
applied the proposal of Hod to investigate the area spectrum of
black holes by virtue of the periodicity of the outgoing wave. It
can also be checked using the adiabatic invariant proposed by
Kunstatter \cite{Kunstatter2003}. For the Kerr black hole, of
course, the modified adiabatic invariant should be adopted \cite
{Vagenas0811}. The obtained area spectrum is valid only for large
values of $n$ since our calculations are based on the semiclassical
approximation. Although the quantum gravity theory has not been
found, it is also meaningful to investigate the quantum correction
to the area spectrum under some heuristic paradigms
\cite{Ali497,Banerjee095}.

According to the quantized area spectrum, we also can get the mass
spectrum. From Eq.(\ref{14}), we find that the mass spectrum takes
the form
\begin{equation}
M=\sqrt{\frac{\hbar n}{2}}; \ n=1,2,\cdots \label{32}
\end{equation}
implying the $n \Longrightarrow n - 1$ transition frequency
\begin{equation}
\omega_0\equiv \frac{dM}{\hbar} = \frac{1}{4M}. \label{33}
\end{equation}
Thus quantum
jumps larger than the minimal produce emission at all frequencies are $\omega=\omega_0\delta n $ with $\delta n=1,2,\cdots $.
Bekensten and Mukhanov \cite{Bekenstein7} noted that this simple spectrum can provide a way to make quantum
gravity effects detectable because the spacing is inversely proportional to the black hole mass over all scales.
For the massive black holes,  all the lines in the spectrum are dim and unobservable, which is similar to the  semiclassical Hawking temperature.
 But there exists always a mass regime
for which the first few uniformly spaced lines should be detectable under optimum circumstances.
It was speculated that the detected mass is the primordial mini-black holes \cite{Bekenstein7}.
 Thus  black hole spectroscopy is also useful to look for the  remains of primordial mini-black holes.

\begin{acknowledgements}
Xiao-Xiong Zeng is grateful to Prof. Elias  C. Vagenas for his
helpful correspondence. This research is supported by the National
Natural Science Foundation of China (Grant Nos. 10773002, 10875012,
11175019). It is also supported by the Fundamental Research Funds
for the Central Universities(Grant No. 105116) and the Team Research Program of Hubei University for Nationalities(Grant No. MY2011T006).
\end{acknowledgements}



\end{document}